# Keep your eye on the tank - A fresh look at an accelerating frame demonstration

David Marasco (Foothill College) & Paolo Marasco (San Jose State University)

A water tank sliding down a frictionless ramp is a classic example of an accelerated reference frame. Along with other accelerated frames, Bartlett[1] analyzed frictionless sliding, and showed that the liquid's surface is parallel to the surface of the ramp, rather than the base. Duzen and coworkers[2] used liquid in an accelerating tank to help students explore motion, noting that the ramp case and riding on a cart accelerated by a hanging mass are equivalent. Munera[3] used the second as a lecture demonstration. In both Duzen *et. al.* and Munera, students viewed accelerating tanks from the lab frame, which may be hard to observe at a distance. As the goal of these demonstrations is to understand the effects of the acceleration in the tank frame, tank-frame observations would be beneficial. Following Krulj and Slisko,[4] a camera was mounted to an accelerated object to make in-frame videos of liquid-based equivalence principle demonstrations. Previous work[2,3] did not quantify the actual acceleration of the tank, use of a SmartCart simplifies this measurement. Kinematics data along with tank-frame video for classroom use are included as supplemental materials.

The experimental setup (*Figure 1*) centered on a plastic fishtank intended for refraction demonstrations.[5] The tank was partially filled with canola oil. Oil was selected over water as its viscosity gives a shorter settling time, which is important given the demonstration's time scale. Following Duzen, a rubber band was stretched across the tank parallel to its base as a level reference. For additional calibration, strips of paper with markings in one-centimeter increments were taped to the tank. The tank was mounted on a kinematics cart,[6] which collected velocity vs. time data, and accelerations were determined from the slopes of these graphs. A camera attached to the cart videoed the surface of the liquid in the tank frame. To improve the visibility of the liquid's surface, white paper was used as a background, with blue painter's tape at 50 centimeter intervals as a visual reference for motion. Experiments were performed on a ramp inclined at 2.4 degrees. Three cases were explored.

In the first, the recording starts prior to the release of the tank from the top of the ramp. This allows students to observe the surface of the liquid is not aligned with the reference rubber band. The tank is released and the surface settles into alignment. Figure 2 shows the liquid's surface when the tank is at rest and under acceleration. Using the reference marks on the edges of the tank, the surface edges move by roughly one centimeter up or down between the at-rest and accelerating conditions. The 2cm increase is across the 46cm width of the tank, which indicates an acceleration of 0.4m/s$^2$ parallel to the ramp's surface (9.8m/s$^2$ * 2cm/46cm). This compares well with 0.38 m/s$^2$, the slope of the velocity vs. time graph and 0.41 m/s$^2$, the value predicted by angle of the ramp (9.8m/s$^2$ * Sin(2.4)).

In the second video the tank is given an initial push up the ramp. While there is some slosh, the surface aligns with the reference rubber band after the tank is released and stays aligned as the tank slows, turns around, and then increases speed down the ramp (direction and speed are cued by the vertical stripes in the background). This indicates that the acceleration does not change. The video is visual evidence for students with the common misconception “acceleration is zero at the peak of flight.” This supports the velocity vs. time graph for this measurement, which shows a constant slope (figs. 3 and 4). The acceleration measurements from the liquid surface alignment and slope of the velocity vs. time graph are again consistent at $0.4 m/s^2$ and $0.39 m/s^2$.

Included in the supplementary materials is a video of the tank in motion, however the liquid’s surface is not aligned with the reference rubber band. The cart has been attached to a hanging mass on a pulley, with the result being no net force on the tank. Travelling at constant velocity, the liquid’s surface aligns perpendicular to gravity. This demonstration dissuades students from concluding that when the liquid is aligned with the ground it means the tank is at rest, and when the liquid is aligned with the ramp it means the tank is in motion.  Instead the video reinforces the role of acceleration.

Supplemental materials:
https://drive.google.com/drive/folders/1BW1725ZB06xahBfYUlr7D08n4SXruKeR?usp=sharing

Figures:

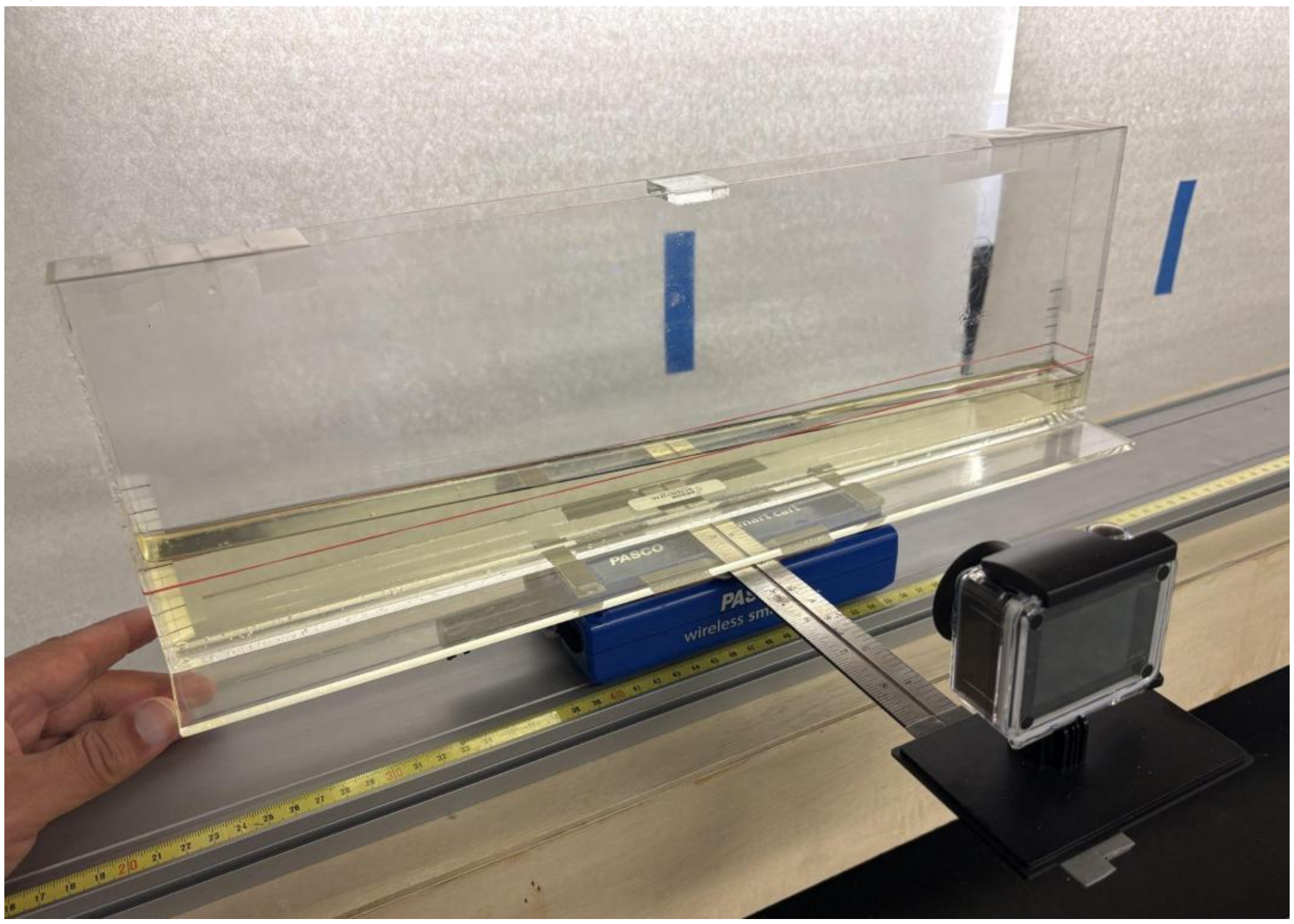

Figure 1 - An oil-filled refraction tank is mounted on a dynamics cart, with a camera to record the liquid surface in the cart frame.

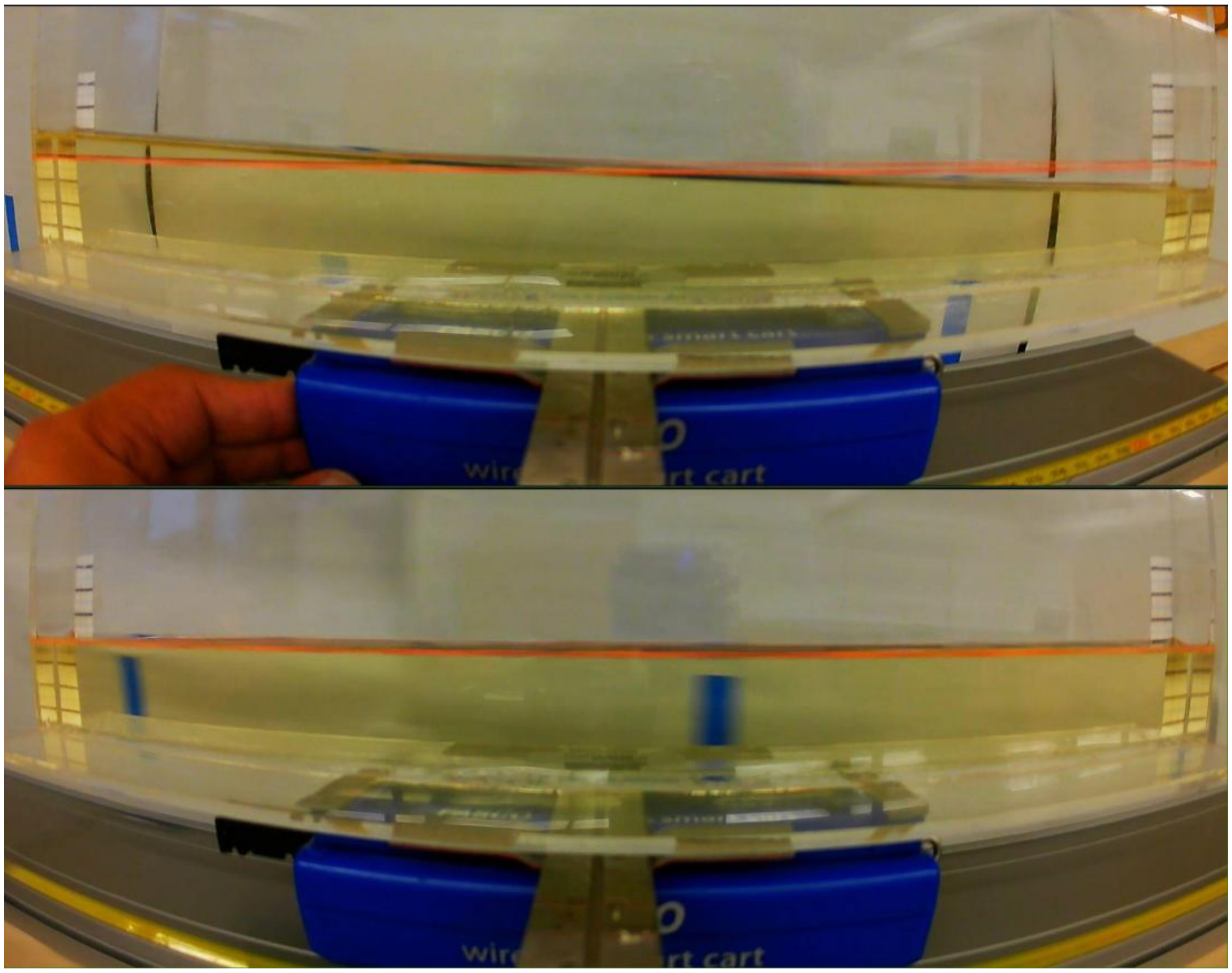

Figure 2 - The top image shows the system prior to release, the liquid’s surface is aligned with the base of the ramp, rather than the rubber band (which is parallel to the ramp surface). The bottom image shows the system as it accelerates down the ramp, with the liquid’s surface aligned with the rubber band.

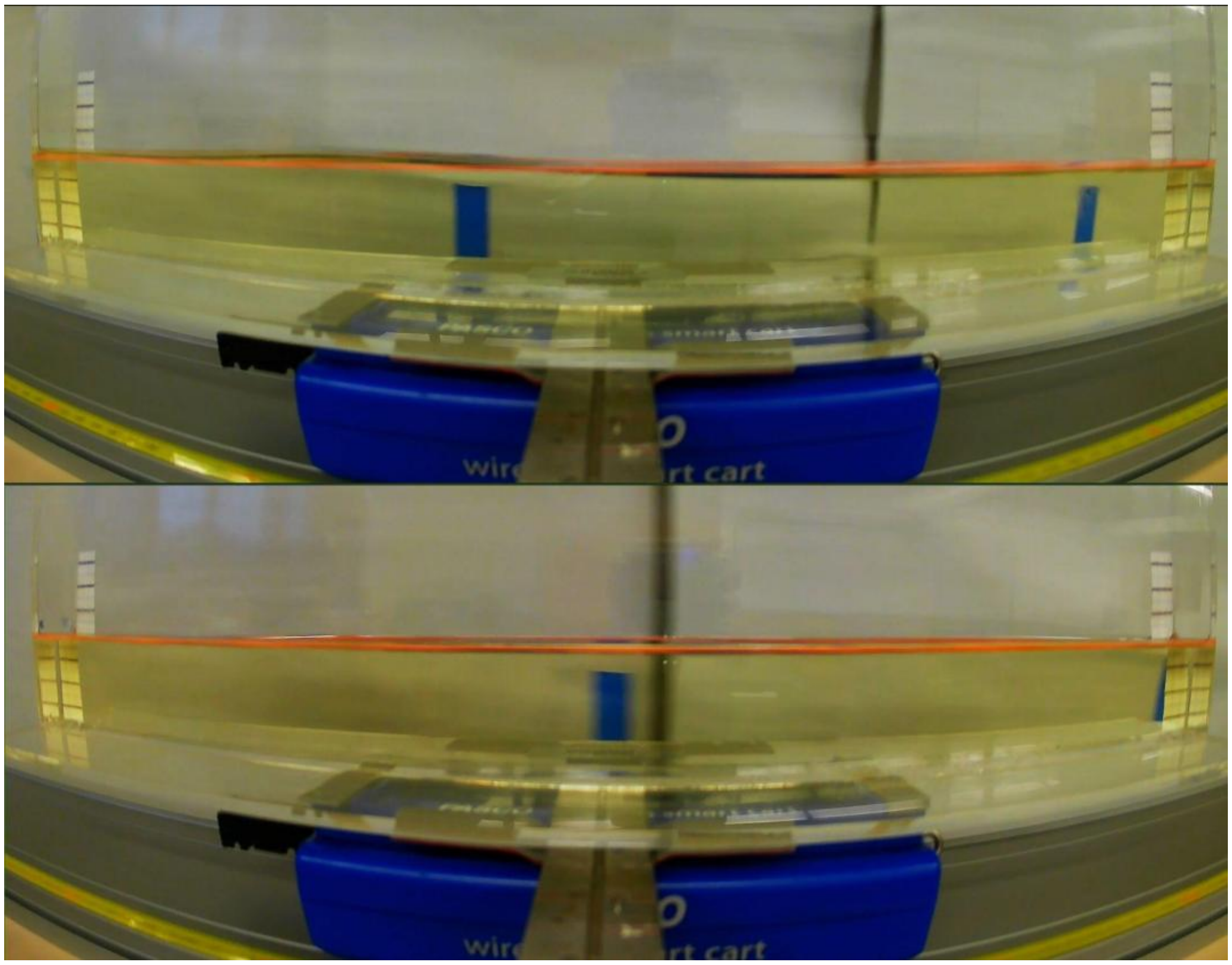

Figure 3 - The liquid's surface aligns with the reference rubber band on the way up the ramp (top) and back down (bottom).

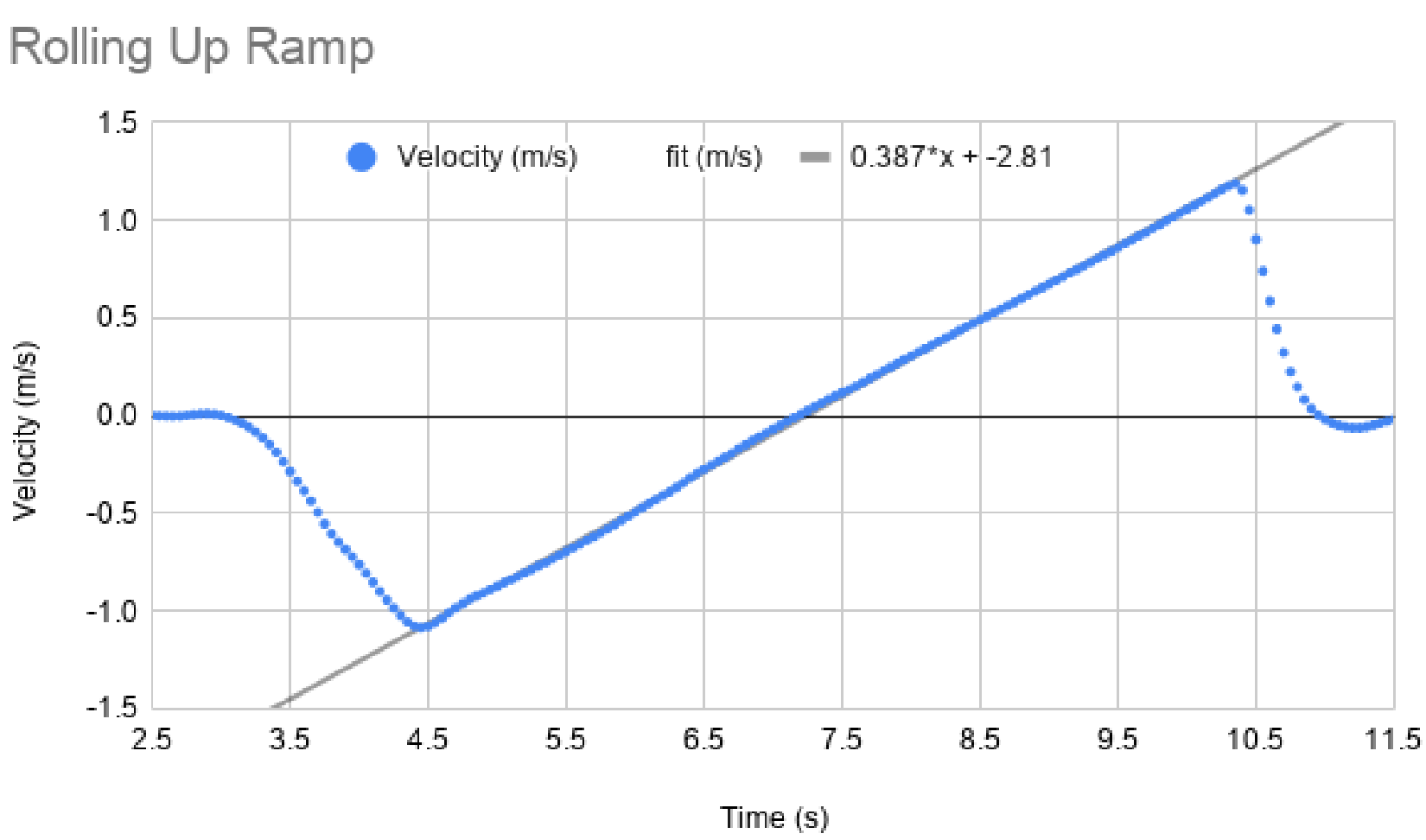

Figure 4 - Acceleration as measured by the slope of the velocity vs. time graph confirms the visual conclusions found in figure 3.